\documentclass[twocolumn,amssymb,nobibnotes,showpacs,superscriptaddress,aps,prd]{revtex4-1}

\usepackage{amsmath,graphicx}

\begin{document}


\title{Squeezed light induced symmetry breaking superradiant phase transition}


\author{C. J. Zhu}
\affiliation{MOE Key Laboratory of Advanced Micro-Structured Materials,
	School of Physics Science and Engineering,Tongji University, Shanghai, China 200092}
\author{L. L. Ping}
\affiliation{MOE Key Laboratory of Advanced Micro-Structured Materials,
	School of Physics Science and Engineering,Tongji University, Shanghai, China 200092}
\author{Y. P. Yang}
\email[Corresponding author:]{yang\_yaping@tongji.edu.cn}
\affiliation{MOE Key Laboratory of Advanced Micro-Structured Materials,
	School of Physics Science and Engineering,Tongji University, Shanghai, China 200092}
\author{G. S. Agarwal}
\email[Corresponding author:]{girish.agarwal@tamu.edu}
\affiliation{Institute for QuantumScience and Engineering, and Department of Biological and Agricultural Engineering Texas A\&M University, College Station, Texas 77843, USA}


\date{\today}

\begin{abstract}
We theoretically investigate the quantum phase transition in the collective systems of qubits in a high quality cavity, which is driven by a squeezed light. We show that the squeezed light induced symmetry breaking can result in quantum phase transition without the ultrastrong coupling requirement. Using the standard mean field theory, we derive the condition of the quantum phase transition. Surprisingly, we show that there exist a tricritical point where the first- and second-order phase transitions meet. With specific atom-cavity coupling strengths, both the first- and second-order phase transition can be controlled by the squeezed light, leading to an optical switching from the normal phase to the superradiant phase by just increasing the squeezed light intensity. The signature of these phase transitions can be observed by detecting the phase space Wigner function distribution with different profiles controlled by the squeezed light intensity. Such superradiant phase transition can be implemented in various quantum systems, including atoms, quantum dots and ions in optical cavities as well as the circuit quantum electrodynamics system.
\end{abstract}


\maketitle
Quantum phase transitions (QPTs) are phase transitions (PT) between two robust stable states caused by a non-thermal parameter that occur at zero temperature ($T=0$ K)~\cite{Sachdev99}. Up to date, the QPTs have received extensive attention both in experiments~\cite{Baumann,feng2015exploring} and theories~\cite{Dimer07,Liu11,Liu13,Zhao17,Baksic14,HEPP73,lambert2004entanglement,lee2004first,buvzek2005instability}. In general, there exist two categories of QPTs: the first-order phase transition and the second-order phase transition. In the second-order phase transitions, this change is abrupt at a critical point. In addition, the quantum fluctuations must diverge at a critical point so that the two kinds of phases with different symmetries can be matched. However, in the first-order phase transitions, these two stable phases can coexist in a critical hysteresis regime, where the quantum fluctuations exhibit a jump across the phase transition boundary.

For cavity quantum electrodynamics (QED) systems, the QPTs are driven by the quantum fluctuations and usually take place for a certain critical atom-cavity coupling strength $g_c$~\cite{larson2017some}. For the coupling strength $g<g_c$, the cavity field is in the vacuum and the atoms are in their ground states, which is recognized as the normal phase (NP) due to the symmetry of system. For $g>g_c$, both the cavity field and the atoms are excited, which represents the superradiant phase (SP) due to the symmetry breaking. In deep strong coupling cases~\cite{casanova2010deep}, the undriven Tavis-Cummings (TC) model with rotating wave approximation has phase transition at the critical coupling strength $g_c\propto\sqrt{\omega_c\omega_A}$ under the non-dissipative regime, where $\omega_{c(A)}$ is the cavity (atom) resonant frequency~\cite{castanos2009coherent,hepp1973equilibrium}. In the dissipative TC model, the phase transition is destroyed and the normal phase replaces the superradiant phase, resulting in the disappearance of the normal-superradiant PT~\cite{soriente2018dissipation}. In ultrastrong coupling cases~\cite{gu2017microwave,armata2017harvesting,niemczyk2010circuit}, where the rotating wave approximation is not valid, the phase transition can also be observed in the undriven Dicke model~\cite{hioe1973phase,bakemeier2012quantum,soriente2018dissipation}, but the critical point is displaced when counter-rotating terms are considered~\cite{Baumann,gutierrez2018dissipative}. It is noted that the normal-superradiant phase transition are also discovered in the Rabi model~\cite{Wang73,ashhab2013superradiance,hwang2015quantum}, Jaynes-Cummings model~\cite{hwang2016quantum} and the driven TC system~\cite{feng2015exploring,zou2013quantum,narducci1978transient,puri1980exact,schneider2002entanglement} by taking the proper thermodynamical limit. Recently, the PT between the symmetric normal state and the state where the NP and SP coexist is proposed in the interpolating Dicke-Tavis-Cummings~\cite{soriente2018dissipation} and Jaynes-Cummings-Rabi~\cite{gutierrez2018dissipative} models, which opens a new direction in this field.


In this letter, we introduce a new direction in the study of quantum phase transitions. We drive the cavity with squeezed light and thus break the inherent U(1) symmetry of the rotating wave Tavis-Cummings model. In general, the TC model with cavity dissipation
%
%
has rotating U(1) symmetry for the transformation $a\rightarrow a{\rm e}^{i\phi}$ and $S_+\rightarrow S_+{\rm e}^{-i\phi}$. In the presence of the squeezed driving field, however, the driving term $G(a^2+a^{\dag2})$ break the symmetry of system, leading to the occurrence of the normal-superradiant PT without the requirement of the ultrastrong coupling.
%
Using the standard mean field approximation and the quantum fluctuation analysis, we find that  the first- and second-order PT coexist in our system and they meet at a critical point. We show that the signature of these superradiant PTs can be observed by detecting the phase space Wigner function distribution, whose profile changes with the squeezed light intensity. Moreover, we show that the PT in our system can be optically controlled, as opposed to many traditional proposals for realizing PT by changing the atom-cavity coupling strength. Using these advantages in our system, an optical switching from normal phase to superradiant PT can be accomplished by increasing the squeezed light intensity. We also notice that there exist several proposals to realize phase transitions by tunning the driving field intensity~\cite{minganti2018spectral,casteels2017critical,carmichael2015breakdown}. As opposite to these proposals, the squeezed field driving will enhance the atom-cavity coupling, leading to the splitting of the normal phase and the implementation of the first-order phase transition. Moreover, with the increase of the driving strength, the system can enter the strong coupling regime and even the ultrastrong coupling regime which is challenging in experiments.

\begin{figure}[htbp]
\includegraphics[width=\linewidth]{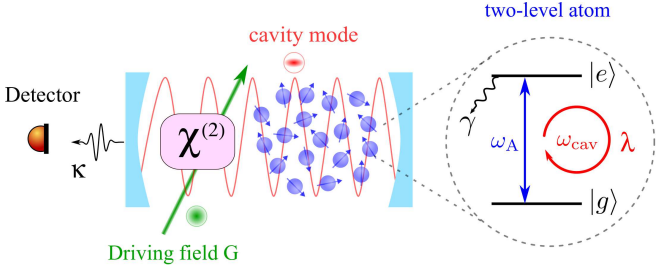}%
\caption{The setup of the atoms cavity-QED system. The cavity mode (red) with angular frequency $\omega_{\rm cav}$ is generated by a strong driving field (green) with angular frequency $2\omega_{\rm sq}$ via the $\chi^{(2)}$ nonlinearity. N identical two-level atoms with resonant frequency $\omega_A$ interact with the cavity mode with atom-cavity coupling strength $\lambda$. The ground (excited) state of the atom is labeled as $|g\rangle$ ($|e\rangle$). Here, $\gamma$ and $\kappa$ are the atomic and cavity decay rates ,respectively.}
~\label{fig:fig1}
\end{figure}
As depicted in Fig.~\ref{fig:fig1}, the system under consideration is a typical Tavis-Cummings model. $N$ identical two-level atoms with resonant frequency $\omega_A$ are trapped inside a single-mode cavity (the mode frequency is $\omega_{\rm cav}$). In our system, the cavity mode is generated through a crystal with second-order nonlinearity~\cite{huang2009normal,leroux2018enhancing}. As shown in Fig.~\ref{fig:fig1}, a strong pump field (green) with angular frequency $2\omega_{\rm sq}$ drives the nonlinear crystal coherently. Then, the generated intracavity field interacts with the atoms with the same coupling strength $\lambda$. Such model can also be realized in many quantum system, such as atoms and ions cavity QED system~\cite{muller2012engineered,bernien2017probing}, optomechanical systems~\cite{ying2014quantum}, polariton condensates~\cite{sieberer2013dynamical} as well as the superconducting qubits~\cite{houck2012chip,feng2015exploring,fitzpatrick2017observation}. Assuming $a^\dag$ and $a$ are the creation and annihilation operators of the cavity mode, respectively, the Hamiltonian of the system is given by
\begin{eqnarray}\label{eq:H}
H&=&\hbar\Delta_C a^\dag a+\hbar\Delta_AS_z+\frac{\hbar\lambda}{\sqrt{N}}(S_+a+S_-a^\dag)\nonumber\\
& &+\hbar G(a^2+a^{\dag2}),
\end{eqnarray}
where $\Delta_C=\omega_{\rm cav}-\omega_{\rm sq}$ and $\Delta_{A}=\omega_{A}-\omega_{\rm sq}$ are the detunings of the cavity and the atoms, respectively. The operators $S_\pm=S_x\pm iS_y$ and $S_\alpha=\sum_{j=1}^N\sigma_\alpha^j/2\ (\alpha=x,y,z)$ are the collective spin operators constituted from the individual Pauli spin operator $\sigma_\alpha^j$ which are used to describe the j-th two-level atom systems. It is noted that the last term on the right-hand side of Eq.~(\ref{eq:H}) represents the driving term by the squeezed field with driving strength $G$. In the absence of the driving term (i.e., $G=0$), such a system has a continuous $U(1)$-symmetry so that the superradiant phase transition will not take place~\cite{soriente2018dissipation}. However, in the presence of the squeezed field (i.e., $G\neq0$), the $U(1)$-symmetry is broken and the superradiant phase transition can be observed, leading to an optical switching from normal phase to superradiant phase by increasing the squeezing strength.

In general, the squeezed field driven properties of the TC model is described by the master equation for density matrix $\rho$ of the system
\begin{eqnarray}\label{eq:master}
\frac{d\rho}{dt}&=&-\frac{i}{\hbar}\left[H,\rho\right]+{\cal L}_{\rm cav}\rho+{\cal L}_{\rm A}\rho,
\end{eqnarray}
where ${\cal L}_{\rm cav}\rho=\kappa(2a\rho a^\dag-a^\dag a\rho-\rho a^\dag a)$ and ${\cal L}_{\rm A}\rho=\gamma(2S_-\rho S_+-S_+S_-\rho-\rho S_+ S_-)/N$ are the dissipations of the cavity and atoms, respectively. In our system, we assume that the atomic decay rate $\gamma$ is much smaller than the cavity decay rate $\kappa$. Although the spontaneous emission may break the phase transition by softening the quantum fluctuations, we set $\gamma=0$ in the following analytical derivations for mathematical simplicity and the conservation of spin is considered, which plays an important role. It is also noted that the dissipation of cavity can also cause the symmetry breaking and alter the phase diagram of the system, which is considered in the following derivations~\cite{soriente2018dissipation}.

To study the phase transition induced by the squeezed field, we begin with the Heisenberg equations of motion for operators, which reads
%
\begin{subequations}\label{eq:HE}
	\begin{align}
&\frac{d a}{dt} = -i\Delta_C a-2iG a^\dag-i\frac{\lambda}{\sqrt{N}}S_x-\frac{\lambda}{\sqrt{N}}S_y-\kappa a,\\
%
%
&\frac{dS_x}{dt} =-\Delta_A S_y +i\frac{\lambda}{\sqrt{N}}S_z(a-a^\dag),\\
&\frac{dS_y}{dt} =\Delta_A S_x -\frac{\lambda}{\sqrt{N}}S_z(a+a^\dag),\\
&\frac{dS_z}{dt} =\frac{\lambda}{\sqrt{N}}S_y(a+a^\dag) -i\frac {\lambda}{\sqrt{N}}S_x(a-a^\dag).
\end{align}
\end{subequations}
The above equations can be linearized by the mean-field approximation by expanding an arbitrary operator $A$ as the form of $A=\langle A \rangle+\delta A$, which is widely used for describing many-body system. Here, $\langle A \rangle$ is the mean value of the operator $A$, and $\delta A$ represents the quantum fluctuation on top of the mean value. For mathematical simplicity, one can rewrite the Eqs.~(\ref{eq:HE}) by using the following definitions $\langle S_x \rangle=NX$, $\langle S_y \rangle=NY$, $\langle S_z \rangle=NZ$ and $\langle a \rangle=\sqrt{N}\alpha$, where $\alpha=\alpha_{\rm Re}+i\alpha_{\rm Im}$.

Under the steady-state approximation and the spin-conservation law $X^2+Y^2+Z^2=1/4$, the Eqs.~(\ref{eq:HE}) can be solved analytically (assuming $\Delta_C=\Delta_A=\Delta$), yielding $X=2\lambda Z\alpha_{\rm Re}/\Delta$ and $Y=-2\lambda Z\alpha_{\rm Im}/\Delta$. To obtain non-trivial solutions of $\alpha_{\rm Re}$ and $\alpha_{\rm Im}$, the constraint on parameter $Z$ is given by
\begin{equation}\label{eq:Z}
	Z_\pm=\frac{\Delta}{2\lambda^2}(-\Delta\pm\sqrt{4G^2-\kappa^2}).
\end{equation}
%
%
%

In Eq.~(\ref{eq:Z}), we assume $\Delta\neq0$, otherwise no phase transition is possible. Obviously, $2G\geq\kappa$ is required to make $Z_\pm$ be real. This condition implies that there exist only trivial solutions (normal phase) for $\alpha_{\rm Re}$ and $\alpha_{\rm Im}$ if $G<\kappa/2$. When $G>\kappa/2$, the superradiant phase may take place, leading to a set of non-trivial solutions. We also note that for $G=0$ and $\kappa\rightarrow0$, the condition for phase transition is $\lambda>|\Delta|$. This is the case experimentally studied by Feng et al~\cite{feng2015exploring}. To obtain stable physical solutions for $\alpha_{\rm Re}$ and $\alpha_{\rm Im}$, we carry out standard stability analysis [see the supplementary material for details]. It is found that only the solutions associated with $Z_{+}$ are stable in our system. Using the spin conservation condition, we get
\begin{subequations}\label{eq:alpha}
	\begin{eqnarray}
		& & \alpha_{\rm Re}=\pm\sqrt{p/\left[q+\frac{q\kappa^2\Delta^2}{(2\lambda^2 Z+\Delta^2-2G\Delta)^2}\right]},\\
		& & \alpha_{\rm Im}=\frac{\kappa\Delta}{2\lambda^2 Z+\Delta^2-2G\Delta}\alpha_{\rm Re},
	\end{eqnarray}
\end{subequations}
where $p=1/4-Z^2$ and $q=4\lambda^2Z^2/\Delta^2$. It is worth noting that the solutions are the opposite of one another, reflecting the $Z_2$ symmetry breaking of system.

\begin{figure}[htbp]
	\includegraphics[width=\linewidth]{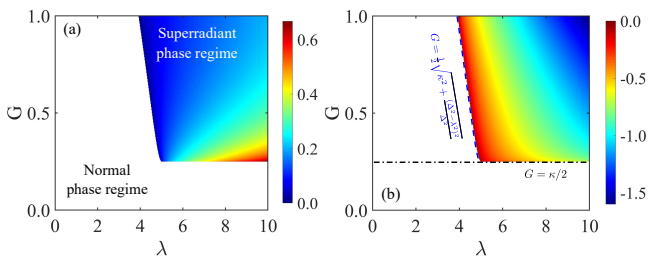}
	\caption{Panels (a) and (b) describe the real and imaginary parts of the expectation value $\langle a\rangle$, respectively, as functions of the atom-cavity coupling strength $\lambda$ and the driving strength $G$ of the squeezed light. The white areas represent $\alpha_{\rm Re}=\alpha_{\rm Im}=0$, corresponding to the normal phase regimes, while the colored areas represent the superradiant phase regimes. In panel (b), the horizontal dash-dotted line and the dashed curve indicate the boundaries from normal phase to superradiant phase. The system parameters are given by $\kappa=0.5$ MHz and $\Delta=5$ MHz.}
\label{fig:fig2}
\end{figure}
In Fig.~2, we plot $\alpha_{\rm Re}$ [panel (a)] and $\alpha_{\rm Im}$ [panel (b)] as functions of the coupling strength $\lambda$ and the driving strength $G$ of the squeezed light, respectively. Here, the system parameters are chosen as $\kappa=0.5$ MHz and $\Delta=5$ MHz. The white areas denote the normal phase regimes, corresponding to $\alpha_{\rm Re}=\alpha_{\rm Im}=0$. The areas with $\alpha_{\rm Re}\neq0$ and $\alpha_{\rm Im}\neq0$ represent the superradiant phase regimes. Obviously, there exist two boundaries for this phase transition indicated in panel (b). One boundary condition satisfies $G=\kappa/2$ (horizontal dash-dotted line), representing that the phase transition occurs only if $G>\kappa/2$. The other boundary condition satisfies $G=\sqrt{\kappa^2+(\Delta^2-\lambda^2)^2/\Delta^2}/2$ (dashed curve) obtained by setting $Z_{+}=-1/2$. In the left-side area of this boundary condition, one can obtain $|Z_+|>1/2$ so that the system is in the normal phase. In the right-side area, stable solutions of $\alpha_{\rm Re}$ and $\alpha_{\rm Im}$ can be obtained, leading to the superradiant phase regime.

To show more characteristics of the superradiant phase transition, we must examine the quantum fluctuations of the operators $a$ and $a^\dag$ on top of the stable mean-field values. Defining the fluctuation operators $\delta a=\sqrt{N}\delta \alpha$ and $\delta S_\beta=N\delta\beta\ (\beta=X,Y,Z)$, the quantum fluctuations can be evaluated by solving the quantum Langevin equations given in the supplementary material.
%
%
For the normal phase, the variance of cavity mode quadrature is given by
\begin{align}
& \langle \delta\alpha^\dag \delta\alpha \rangle\approx\frac{2G^2[(2\Delta^2-\lambda^2)+\Delta^2(\kappa^2-4G^2+\lambda^2)]}{(\kappa^2-4G^2+4\Delta^2)[(\Delta^2-\lambda^2)+\Delta^2(\kappa^2-4G^2)]}.\nonumber
\end{align}
However, the explicit expression of the variance of cavity mode quadrature for the superradiant phase is too lengthy to be given.

\begin{figure}[htbp]
	\includegraphics[width=\linewidth]{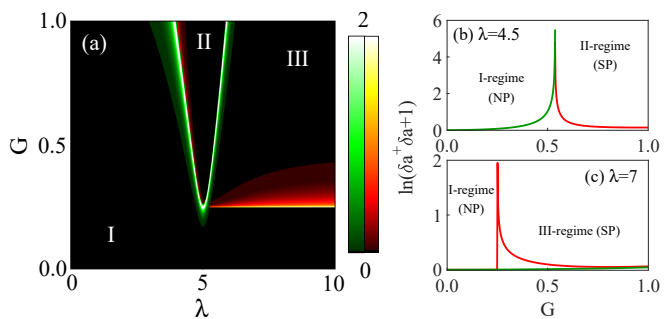}
	\caption{(a) The photon number fluctuations calculated on top of the stable mean-field solutions versus the atom-cavity coupling strength and the  driving strength of squeezed field. Green and red refer to fluctuations on top of the normal phase and the superradiant phase, respectively. In panel (a), each regime is marked by the number of stable physical solutions. Panels (b) and (c) show the photon number fluctuations at (a) $\lambda=4.5$ MHz and (b) $\lambda=7$ MHz, respectively. }
\label{fig:fig3}
\end{figure}
In Fig. 3(a), we show the expected value of the photon number fluctuations $\langle \delta a^\dag \delta a\rangle$ evaluated based on the mean-field solutions. For clarity, ${\rm ln}(\langle \delta a^\dag \delta a\rangle+1)$ is plotted as functions of the coupling strength $\lambda$ and the driving strength $G$. Here, the green and red correspond to the fluctuations of the normal phase and the superradiant phase, repsectively. Since the normal phase splitting induced by the driving term, it is clear to see that there exist three regime for which the photon fluctuations vanish, marked by the number of coexisting solutions. From regime I to regime II, the photon fluctuations diverge continuously across both sides of the transition, known as the continuous second-order transition which morph into two discontinuous first-order transitions [see panel (b)]. From regime I to regime III, however, one can observe discontinuous first-order transition between the normal phase and the superradiant phase accompanying with the coexistence regime [see panel (c)].

\begin{figure}[htbp]
	\includegraphics[width=\linewidth]{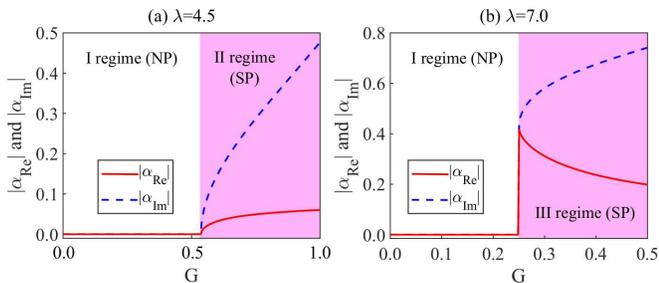}
	\caption{Optical switching from normal phase to superradiant phase by increasing the driving strength $G$. The red solid and blue dashed curves denote $|\alpha_{\rm Re}|$ and $|\alpha_{\rm Im}|$, respectively. The coupling strengths are chosen as $\lambda=4.5$ MHz and $\lambda=7$ MHz in panel (a) and (b), respectively.}
\label{fig:fig4}
\end{figure}
In sight of these features demonstrated above, optical switching from the normal phase to the superradiant phase can be accomplished by increasing the driving strength of the squeezed light. In Fig. 4, we demonstrate two kinds of optical switching operations by taking $\lambda=4.5$ MHz (a) and $\lambda=7$ MHz (b), respectively. One is the switching from normal phase to superradiant phase without coexisting solutions [see panel (a)]. The other is the switching from normal phase to superradiant phase with coexisting solutions [see panel (b)]. As shown in Fig.~\ref{fig:fig4}(a), the values of $\alpha_{\rm Re}$ and $\alpha_{\rm Im}$ change continuously at $G>\kappa$, representing an expected second-order transition. However, there exist an abrupt change for the values of $\alpha_{\rm Re}$ and $\alpha_{\rm Im}$ at $G=\kappa/2$, which is a witness for the first-order phase transition.

\begin{figure}[htbp]
	\includegraphics[width=\linewidth]{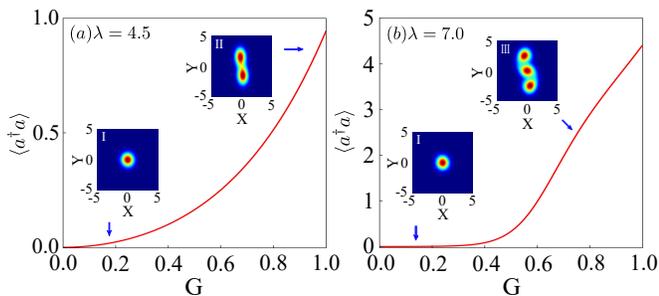}
	\caption{Mean photon number $\langle a^\dag a\rangle$ calculated by directly solving the master equation is plotted as a function of the driving strength $G$ for (a) $\lambda=4.5$ MHz and (b) $\lambda=7$ MHz, respectively. The inserted figures are the density plot of the Wigner functions in regime I, II, and III.}
\label{fig:fig5}
\end{figure}
To verify these analytical solutions, we carry out full numerical calculation by directing solving the master equations. Here, we only consider four two-level atoms and the accuracy of our numerical calculations is limited by the finite size of the Hilbert space spanned by 50 photon states. We must point out that the number of atoms is enough to exhibit all the properties of the phase transition process. In our numerical simulation, we also consider the atomic decay by taking $\gamma=0.001$ MHz, which is valid for the Rdyberg sates and metastable states coupled by two-photon process. As shown in Fig. 5, the mean photon number grows continuously for $\lambda=4.5$ MHz, while an abrupt increase in the mean photon number can be observed for $\lambda=7$ MHz, which agrees well with our analytical results.

Although the mean photon number changes as expected, the superradiant phase transition can't be assessed by just measuring the mean photon number. Thus, we calculate the Wigner function to show the phase characteristics of the cavity field. For small drive strength, the system is in the normal phase regime and the Wigner function is that of the vacuum state, which has Gaussian distribution (regime I). When the system enters into the superradiant phase regime, the Wigner function varies significantly. For $\lambda=4.5$ MHz, there exist two peaks in the Wigner function distribution and the shapes of these two peaks are compressed in the X-direction. This double-peaked Wigner function demonstrate the squeezed superradiant phase (regime II) clearly. For $\lambda=7$ MHz, there exist three peaks in the Wigner function, which show the mixture of the normal phase and the superradiant phase (regime III).

In summary, we have studied the quantum phase transitions induced by a squeezed light in atoms cavity QED system. Using the standard mean field approximation and quantum fluctuation analysis, we linearize the equations for operators and find the critical point for realizing the superradiant phase transition. Increasing the squeezed field intensity, moreover, an optical switching from the normal phase to superradiant phase can be accomplished. By analyzing the quantum fluctuation and detecting the phase space Wigner function distribution, we show that the normal phase undergoes a splitting with the increase of the driving strength. As a result, not only the first-order phase transition but also the second-order phase transition can be achieved if a specific atoms-cavity coupling strength is chosen. For the second-order phase transition, we show that the system is changed from the normal phase with classical mode to the superradiant phase with two squeezed modes continuously. For the first-order phase transition, however, a sudden change from the normal phase to the superradiant phase with coexisting modes can be accomplished in this system. These results demonstrated in this paper can be used to optically control the phase transitions and the feasibility of our scheme has direct implications for various quantum systems, including the atoms, quantum dots and ions cavity QED systems as well as the circuit QED system.

\appendix
\noindent{\bf Supplementary material} - Here, we provide more detail information about the stability analysis, quantum noise calculation, the calculation of the Wigner function and more details of numerical simulation, respectively. 

\section{Stability analysis}
In general, the solutions of equation (3) aren't all stable. Therefore, it is important to carry out the stability analysis to check the stability of these solutions~\cite{PhysRevA.85.013817}. For any solution of the linear ordinary differential equation $\dot{f}(t)=Af(t)$, if all the eigenvalues of the matrix $A$ have negative real part, this solution is stable, otherwise this solution is unstable. For equation (3), the coefficient matrix $A$ is written as
%
\begin{eqnarray}
A=
\left(
\begin{array}{ccccc}\nonumber
-\kappa    & \Delta-2G &       0     & -\sqrt{2}\lambda &   0\\
-\Delta-2G &   -\kappa   & -\sqrt{2}\lambda    & 0        &   0\\
0        &-\sqrt{2}\lambda Z  &0            & -\Delta &-2\lambda\alpha_{Im}\\
-\sqrt{2}\lambda Z &     0        &\Delta     & 0         &-2\lambda\alpha_{Re}\\
\sqrt{2}\lambda Y & \sqrt{2}\lambda X   & 2\lambda\alpha_{Im} & 2\lambda\alpha_{Re}&0\\
\end{array}.
\right)
\end{eqnarray}
%
%
\setcounter{figure}{0}
\makeatletter 
\renewcommand{\thefigure}{S\@arabic\c@figure}
\makeatother
\begin{figure}[!htp]
	\centering
	\includegraphics[width=\linewidth]{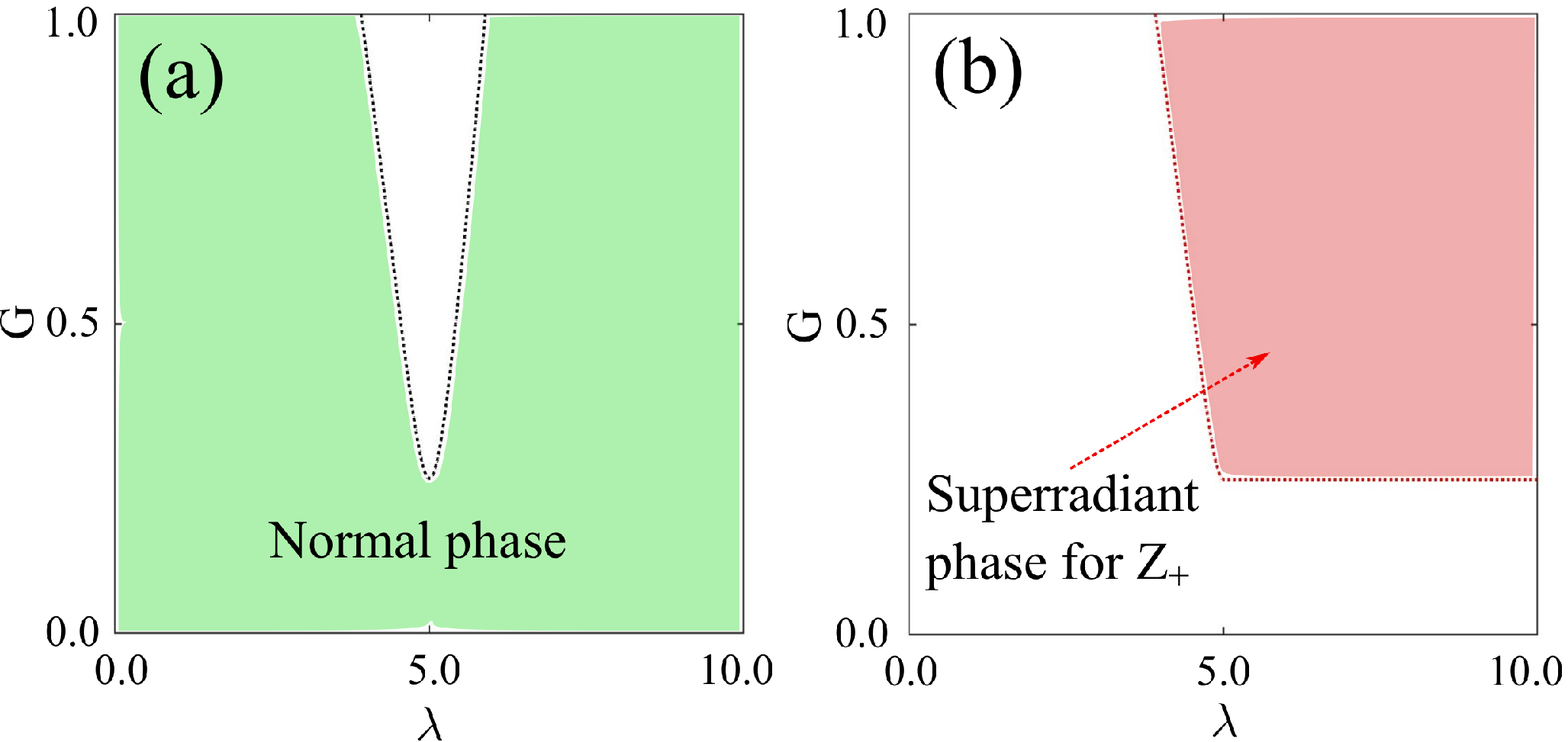}
	\caption{\label{fig:stability} Stable regime of the (a) normal phase and (b) superradiation phase. The white areas indicate the unstable regimes of the normal phase and superradiant phase, respectively.}
\end{figure}
By using the standard stability analysis, we show that the solutions for the normal phase are stable. Correspondingly, the stable regime of the normal phase is illustrated in Fig.~\ref{fig:stability}(a). For the superradiant phase, however, the solutions constrained by $Z_+$ are stable, while the solutions constrained by $Z_-$ are physical unstable. The stability regime of the superradiant phase is shown in Fig.~\ref{fig:stability}(b).

\section{Quantum fluctuations}
According to the standard quantum optical procedure of studying the quantum fluctuations~\cite{agarwal2012quantum}, we define the fluctuation operators as $\delta a=\sqrt{N}\delta \alpha$ and $\delta S_\beta=N\delta\beta\ (\beta=X,Y,Z)$ and linearize the nonlinear equations [Eqs.~(3)]. Then, the linearized quantum Langevin equations for the fluctuation operators take the form
%
%
\begin{align}\label{eq:delta}
\dot{\delta\alpha}=& -i\Delta\delta\alpha -2iG\delta\alpha^\dag-i\lambda\delta X -\lambda\delta Y-\kappa\delta\alpha\nonumber\\
& +\sqrt{2\kappa}\delta\alpha_{\rm in},\tag{S1a}\\
%
%
\dot{\delta X}=& -\Delta\delta Y-2\lambda\alpha_{\rm Im}\delta Z+i\lambda Z(\delta\alpha-\delta\alpha^\dag),\tag{S1b}\\
\dot{\delta Y}=& \Delta\delta X-2\lambda\alpha_{\rm Re}\delta Z-\lambda Z(\delta\alpha+\delta\alpha^\dag),\tag{S1c}\\
\dot{\delta Z}=& 2\lambda\alpha_{\rm Re}\delta Y+\lambda Y(\delta\alpha+\delta\alpha^\dag)+2\lambda\alpha_{\rm Im}\delta X\nonumber\\
& -i\lambda X(\delta\alpha-\delta\alpha^\dag),\tag{S1d}
\end{align}
%
where $\delta\alpha_{\rm in}$ is the fluctuation on top of the zero mean value operator $a_{\rm in}(t)$, describing the vacuum input noise operator of the squeezed light, and satisfying the commutation relation $[a_{\rm in}(t),a_{\rm in}^\dag(t')]=\delta(t-t')$ and the Markovian correlation function $\langle a_{\rm in}(t)a_{\rm in}^\dag(t') \rangle=\delta(t-t')$. 

Introducing the amplitude and phase fluctuation quadratures of the cavity field $\delta Q=(\delta\alpha+\delta\alpha^\dag)/\sqrt{2}$, $\delta P=i(\delta\alpha^\dag-\delta\alpha)/\sqrt{2}$, and the input noise quadratures $\delta Q_{\rm in}=(\delta\alpha_{\rm in}+\delta\alpha_{\rm in}^\dag)/\sqrt{2}$, $\delta P_{\rm in}=i(\delta\alpha_{\rm in}^\dag-\delta\alpha_{\rm in})/\sqrt{2}$, the Eqs.~(S1) can be written in the following compact matrix form
\begin{align}\label{eq:f}
&\dot{f}(t)=Af(t)+\eta(t),\tag{S2}
\end{align}
where $f(t)=[\delta Q,\delta P,\delta X,\delta Y,\delta Z]^T$ is the column vector of the fluctuation, and $\eta(t)=[\sqrt{2\kappa}\delta Q_{\rm in},\sqrt{2\kappa}\delta P_{\rm in},0,0,0]^T$ is the column vector of the noise sources. 

To explore the variances of cavity mode quadratures, we define the time-dependent covariance matrix $V(t)$ with $V_{ij}(t)=\langle f_i(t)f_j(t')+f_j(t')f_i(t)\rangle/2\ (i,j=1-5)$. Using the Eq.~(\ref{eq:f}), one can easily find the solutions of $V(\infty)$ by solving the so-called Lyapunov equation $AV+VA^T=-D$, where $D$ is the diffusion matrix, defined as $D_{ij}\delta(t-t')=\langle \eta_i(t)\eta_j(t')+\eta_j(t')\eta_i(t)\rangle/2\ (i,j=1-5)$. The diagonal elements of the matrix $V$ correspond to the variance of quadratures. In our model, the variance of cavity mode quadrature is given by $\langle \delta\alpha^\dag \delta\alpha \rangle=[(V_{11}+V_{22})-1]/2$.

\section{The Wigner Function}
Wigner function is an important phase-space quasi-probability distribution function proposed by  Wigner~\cite{PhysRev.40.749} in 1932. To date, the Wigner function has been proved to be useful in quantum optics, especially in the representation and visualization of non-classical fields~\cite{agarwal2012quantum}. As known to all, the Wigner function is defined as
\begin{align}\label{w1}
&W(\alpha)=\frac{1}{\pi^2}\int d^2\beta C_W(\beta) e^{\beta^*\alpha-\beta\alpha^*},\tag{S3}
\end{align}
where the characteristic function $C_W(\beta)={\rm Tr}(\rho D(\beta))$ with $\rho$ being the density matrix and $D(\beta)=\exp(\beta a^\dag -\beta^* a)$ being the displacement operator, respectively. 

\begin{figure}[!htp]
	\centering
	\includegraphics[width=\linewidth]{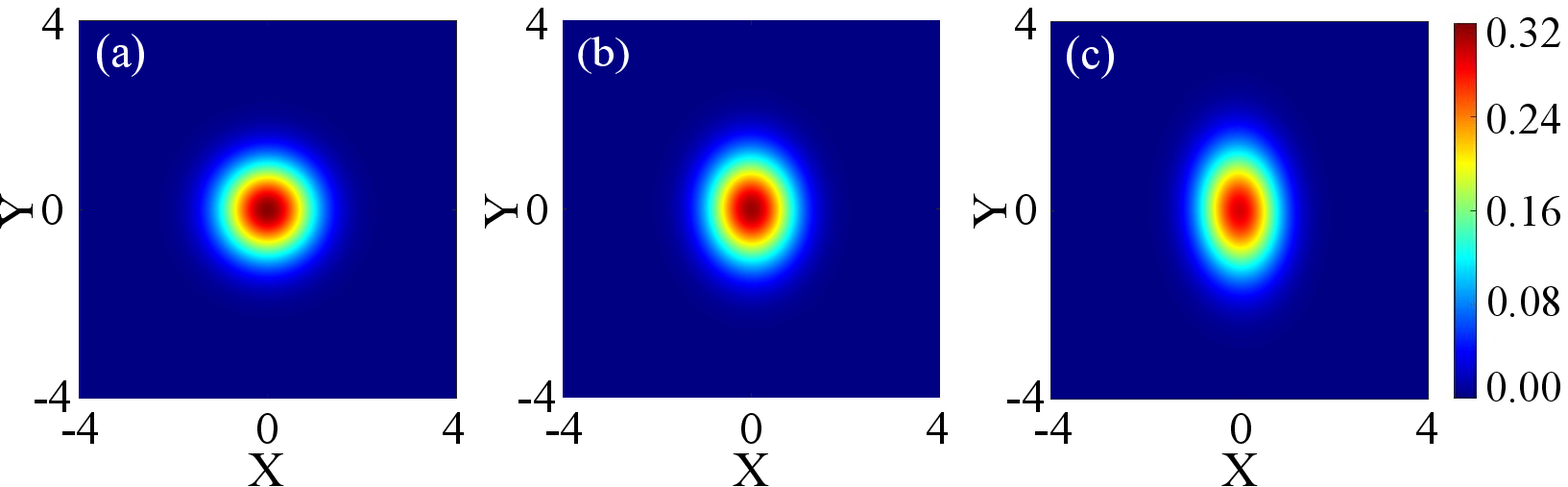}
	\caption{The Wigner functions in an empty cavity with driving term $G(a^{\dag2}+a^2)$. Here, the driving field intensity is chosen as (a) $G=0$, (b) $G=\kappa$ (c) $G=2\kappa$, respectively.~\label{fig:wigner} 
	}
\end{figure}
To understand the characteristics of the Wigner function in superradiant phase, we first consider the case of an empty cavity driven by a squeezed field. In Fig.~\ref{fig:wigner}, we show the Wigner function changing with the driving field intensity. Here, the horizontal and vertical axises correspond to the real and imaginary parts of the parameter $\alpha$, respectively. Clearly, the Wigner function has a Gaussian profile if $G=0$, corresponding to a vacuum state. As the driving field strength increases, the profile of the Wigner function is compressed in $X$ axis, representing that the cavity field is in a squeezed state. 

\section{Numerical simulation}
\begin{figure}[htbp]
	\includegraphics[width=\linewidth]{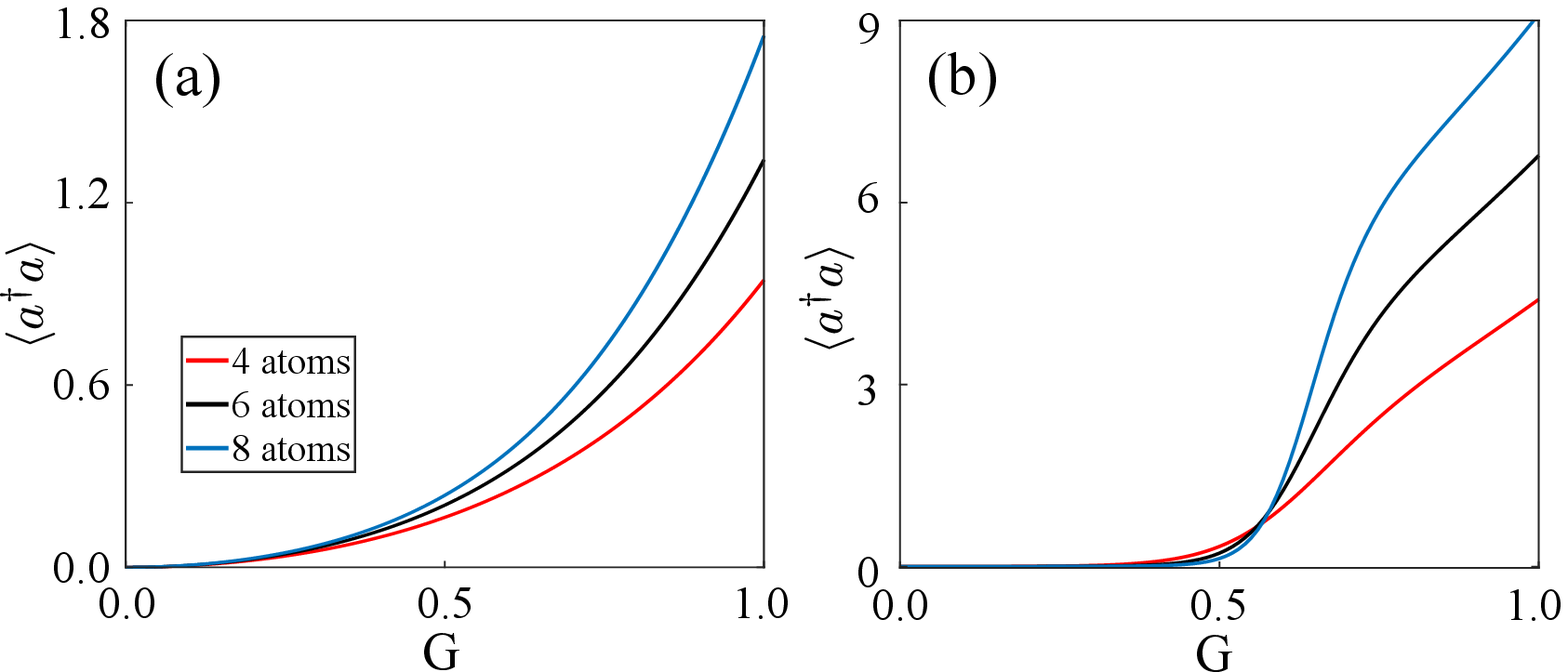}
	\caption{Plot of the mean photon number versus the driving field strength $G$ with the atom-cavity coupling strength (a) $\lambda=4.5$ MHz and (b) $\lambda=7$ MHz, respectively. Here, the number of atoms is chosen as four (red), six (black) and eight (blue), respectively.}\label{fig:G}
\end{figure}
In our numerical simulations, we consider the case of four atoms in the cavity. Although the results agree well with the analytical results, which is obtained by assuming the number of atoms is infinity, it is necessary to examine the multiatoms case. In Fig.~\ref{fig:G}, we plot the mean photon number as a function of the driving field strength $G$ with the atom-cavity coupling strength (a) $\lambda=4.5$ MHz and (b) $\lambda=7$ MHz, respectively. Here, the number of atoms in the cavity is chosen as $N=4$ (red curves), $N=6$ (black curves) and $N=8$ (blue curves), respectively. To make sure the convergence of the numerical calculation, we expand the photon number space up to $N_{\rm ph}=50$. It is clear to see that the profiles of the mean photon number are similar, although the amplitude of the mean photon number enhances with the increase of the number of atoms. 

%
\begin{figure}[htbp]
	\includegraphics[width=\linewidth]{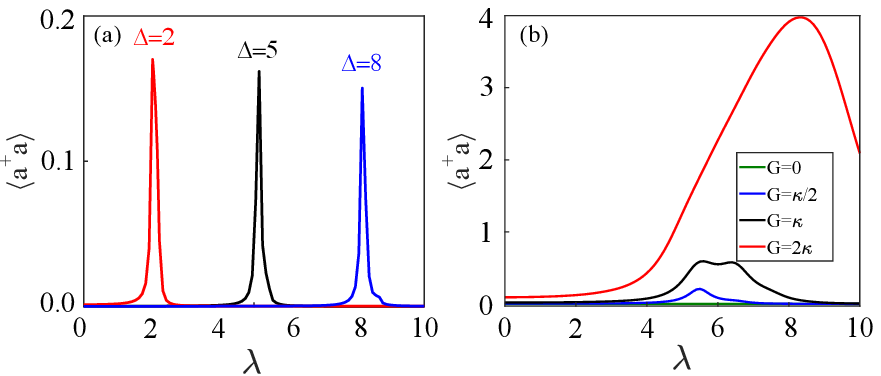}
	\caption{Plot of the mean photon number versus the coupling strength $\lambda$ with separate detuning (a) and driving strength (b). }\label{Fig:lambda}
\end{figure}
Assuming $G=\kappa/2$, one can easily obtain $\lambda=\Delta$ by taking $Z_+=-1/2$, which is the critical point for the quantum phase transition. Thus, the position of the critical point can be changed by adjusting the detuning $\Delta$. To show this feature, we choose $\kappa=0.5$ MHz and plot the mean photon number as a function of the atom-cavity coupling strength in Fig.~\ref{Fig:lambda}(a). Here, the driving field strength $G=\kappa/2$ and the detuning is chosen as $\Delta=2$ MHz (red curve), $\Delta=5$ MHz (black curve) and $\Delta=8$ MHz (blue curve), respectively. Obviously, there exist a single peak at $\lambda=\Delta$, resulting from the quantum fluctuations of the normal phase. The amplitude of this peak decreases slightly as the detuning increases. In Fig.~\ref{Fig:lambda}(b), we set $\Delta=5$ MHz and plot the mean photon number with separate driving strength. With the increase of $G$, the regime of $\lambda>\Delta$ enters into the state of superradiant phase, leading to a boost of photon number. However, in the regime of $\lambda<\Delta$, the system is in the normal phase, and then the mean photon number is close to zero. 

\begin{acknowledgments}
We acknowledge discussions with Dr.~Ricardo Gutierrez Jauregui and Dr.~Jie Li. CJZ and YPY thank the support from the National Key Basic Research Special Foundation (Grant No.2016YFA0302800); the Shanghai Science and Technology Committee (Grants No.18JC1410900) and the National Nature Science Foundation (Grant No.11774262). GSA thanks the support of Air Force Office of  scientific Research (Award No. FA-9550-18-1-0141).
\end{acknowledgments}


\bibliography{Refs}

\begin{thebibliography}{45}%
\makeatletter
\providecommand \@ifxundefined [1]{%
 \@ifx{#1\undefined}
}%
\providecommand \@ifnum [1]{%
 \ifnum #1\expandafter \@firstoftwo
 \else \expandafter \@secondoftwo
 \fi
}%
\providecommand \@ifx [1]{%
 \ifx #1\expandafter \@firstoftwo
 \else \expandafter \@secondoftwo
 \fi
}%
\providecommand \natexlab [1]{#1}%
\providecommand \enquote  [1]{``#1''}%
\providecommand \bibnamefont  [1]{#1}%
\providecommand \bibfnamefont [1]{#1}%
\providecommand \citenamefont [1]{#1}%
\providecommand \href@noop [0]{\@secondoftwo}%
\providecommand \href [0]{\begingroup \@sanitize@url \@href}%
\providecommand \@href[1]{\@@startlink{#1}\@@href}%
\providecommand \@@href[1]{\endgroup#1\@@endlink}%
\providecommand \@sanitize@url [0]{\catcode `\\12\catcode `\$12\catcode
  `\&12\catcode `\#12\catcode `\^12\catcode `\_12\catcode `\%12\relax}%
\providecommand \@@startlink[1]{}%
\providecommand \@@endlink[0]{}%
\providecommand \url  [0]{\begingroup\@sanitize@url \@url }%
\providecommand \@url [1]{\endgroup\@href {#1}{\urlprefix }}%
\providecommand \urlprefix  [0]{URL }%
\providecommand \Eprint [0]{\href }%
\providecommand \doibase [0]{http://dx.doi.org/}%
\providecommand \selectlanguage [0]{\@gobble}%
\providecommand \bibinfo  [0]{\@secondoftwo}%
\providecommand \bibfield  [0]{\@secondoftwo}%
\providecommand \translation [1]{[#1]}%
\providecommand \BibitemOpen [0]{}%
\providecommand \bibitemStop [0]{}%
\providecommand \bibitemNoStop [0]{.\EOS\space}%
\providecommand \EOS [0]{\spacefactor3000\relax}%
\providecommand \BibitemShut  [1]{\csname bibitem#1\endcsname}%
\let\auto@bib@innerbib\@empty
\bibitem [{\citenamefont {Sachdev}(1999)}]{Sachdev99}%
  \BibitemOpen
  \bibfield  {author} {\bibinfo {author} {\bibfnamefont {S.}~\bibnamefont
  {Sachdev}},\ }\href
  {https://onlinelibrary.wiley.com/doi/abs/10.1002/9780470022184.hmm108} {\emph
  {\bibinfo {title} {Quantum Phase Transitions}}}\ (\bibinfo  {publisher}
  {Cambridge University Press},\ \bibinfo {year} {1999})\BibitemShut {NoStop}%
\bibitem [{\citenamefont {Baumann}\ \emph {et~al.}(2010)\citenamefont
  {Baumann}, \citenamefont {Guerlin}, \citenamefont {Brennecke},\ and\
  \citenamefont {Esslinger}}]{Baumann}%
  \BibitemOpen
  \bibfield  {author} {\bibinfo {author} {\bibfnamefont {K.}~\bibnamefont
  {Baumann}}, \bibinfo {author} {\bibfnamefont {C.}~\bibnamefont {Guerlin}},
  \bibinfo {author} {\bibfnamefont {F.}~\bibnamefont {Brennecke}}, \ and\
  \bibinfo {author} {\bibfnamefont {T.}~\bibnamefont {Esslinger}},\ }\bibfield
  {title} {\enquote {\bibinfo {title} {Dicke quantum phase transition with a
  superfluid gas in an optical cavity},}\ }\href {\doibase 10.1038/nature09009}
  {\bibfield  {journal} {\bibinfo  {journal} {Nature}\ }\textbf {\bibinfo
  {volume} {464}},\ \bibinfo {pages} {1301--1306} (\bibinfo {year}
  {2010})}\BibitemShut {NoStop}%
\bibitem [{\citenamefont {Feng}\ \emph {et~al.}(2015)\citenamefont {Feng},
  \citenamefont {Zhong}, \citenamefont {Liu}, \citenamefont {Yan},
  \citenamefont {Yang}, \citenamefont {Twamley},\ and\ \citenamefont
  {Wang}}]{feng2015exploring}%
  \BibitemOpen
  \bibfield  {author} {\bibinfo {author} {\bibfnamefont {M.}~\bibnamefont
  {Feng}}, \bibinfo {author} {\bibfnamefont {Y.~P.}\ \bibnamefont {Zhong}},
  \bibinfo {author} {\bibfnamefont {T.}~\bibnamefont {Liu}}, \bibinfo {author}
  {\bibfnamefont {L.~L.}\ \bibnamefont {Yan}}, \bibinfo {author} {\bibfnamefont
  {W.~L.}\ \bibnamefont {Yang}}, \bibinfo {author} {\bibfnamefont
  {J.}~\bibnamefont {Twamley}}, \ and\ \bibinfo {author} {\bibfnamefont
  {H.}~\bibnamefont {Wang}},\ }\bibfield  {title} {\enquote {\bibinfo {title}
  {Exploring the quantum critical behaviour in a driven tavis--cummings
  circuit},}\ }\href {\doibase 10.1038/ncomms8111} {\bibfield  {journal}
  {\bibinfo  {journal} {Nat. Commun.}\ }\textbf {\bibinfo {volume} {6}},\
  \bibinfo {pages} {7111} (\bibinfo {year} {2015})}\BibitemShut {NoStop}%
\bibitem [{\citenamefont {Dimer}\ \emph {et~al.}(2007)\citenamefont {Dimer},
  \citenamefont {Estienne}, \citenamefont {Parkins},\ and\ \citenamefont
  {Carmichael}}]{Dimer07}%
  \BibitemOpen
  \bibfield  {author} {\bibinfo {author} {\bibfnamefont {F.}~\bibnamefont
  {Dimer}}, \bibinfo {author} {\bibfnamefont {B.}~\bibnamefont {Estienne}},
  \bibinfo {author} {\bibfnamefont {A.~S.}\ \bibnamefont {Parkins}}, \ and\
  \bibinfo {author} {\bibfnamefont {H.~J.}\ \bibnamefont {Carmichael}},\
  }\bibfield  {title} {\enquote {\bibinfo {title} {Proposed realization of the
  dicke-model quantum phase transition in an optical cavity qed system},}\
  }\href {\doibase 10.1103/PhysRevA.75.013804} {\bibfield  {journal} {\bibinfo
  {journal} {Phys. Rev. A}\ }\textbf {\bibinfo {volume} {75}},\ \bibinfo
  {pages} {013804} (\bibinfo {year} {2007})}\BibitemShut {NoStop}%
\bibitem [{\citenamefont {Liu}\ \emph {et~al.}(2011)\citenamefont {Liu},
  \citenamefont {Lian}, \citenamefont {Ma}, \citenamefont {Xiao}, \citenamefont
  {Chen}, \citenamefont {Liang},\ and\ \citenamefont {Jia}}]{Liu11}%
  \BibitemOpen
  \bibfield  {author} {\bibinfo {author} {\bibfnamefont {N.}~\bibnamefont
  {Liu}}, \bibinfo {author} {\bibfnamefont {J.~L.}\ \bibnamefont {Lian}},
  \bibinfo {author} {\bibfnamefont {J.}~\bibnamefont {Ma}}, \bibinfo {author}
  {\bibfnamefont {L.~T.}\ \bibnamefont {Xiao}}, \bibinfo {author}
  {\bibfnamefont {G.}~\bibnamefont {Chen}}, \bibinfo {author} {\bibfnamefont
  {J.~Q.}\ \bibnamefont {Liang}}, \ and\ \bibinfo {author} {\bibfnamefont
  {S.~T.}\ \bibnamefont {Jia}},\ }\bibfield  {title} {\enquote {\bibinfo
  {title} {Light-shift-induced quantum phase transitions of a bose-einstein
  condensate in an optical cavity},}\ }\href {\doibase
  10.1103/PhysRevA.83.033601} {\bibfield  {journal} {\bibinfo  {journal} {Phys.
  Rev. A}\ }\textbf {\bibinfo {volume} {83}},\ \bibinfo {pages} {033601}
  (\bibinfo {year} {2011})}\BibitemShut {NoStop}%
\bibitem [{\citenamefont {Liu}\ \emph {et~al.}(2013)\citenamefont {Liu},
  \citenamefont {Li},\ and\ \citenamefont {Liang}}]{Liu13}%
  \BibitemOpen
  \bibfield  {author} {\bibinfo {author} {\bibfnamefont {N.}~\bibnamefont
  {Liu}}, \bibinfo {author} {\bibfnamefont {J.~D.}\ \bibnamefont {Li}}, \ and\
  \bibinfo {author} {\bibfnamefont {J.~Q.}\ \bibnamefont {Liang}},\ }\bibfield
  {title} {\enquote {\bibinfo {title} {Nonequilibrium quantum phase transition
  of bose-einstein condensates in an optical cavity},}\ }\href {\doibase
  10.1103/PhysRevA.87.053623} {\bibfield  {journal} {\bibinfo  {journal} {Phys.
  Rev. A}\ }\textbf {\bibinfo {volume} {87}},\ \bibinfo {pages} {053623}
  (\bibinfo {year} {2013})}\BibitemShut {NoStop}%
\bibitem [{\citenamefont {Zhao}\ \emph {et~al.}(2017)\citenamefont {Zhao},
  \citenamefont {Liu}, \citenamefont {Bai},\ and\ \citenamefont
  {Liang}}]{Zhao17}%
  \BibitemOpen
  \bibfield  {author} {\bibinfo {author} {\bibfnamefont {X.~Q.}\ \bibnamefont
  {Zhao}}, \bibinfo {author} {\bibfnamefont {N.}~\bibnamefont {Liu}}, \bibinfo
  {author} {\bibfnamefont {X.~M.}\ \bibnamefont {Bai}}, \ and\ \bibinfo
  {author} {\bibfnamefont {J.~Q.}\ \bibnamefont {Liang}},\ }\bibfield  {title}
  {\enquote {\bibinfo {title} {Dicke phase transition and collapse of
  superradiant phase in optomechanical cavity with arbitrary number of
  atoms},}\ }\href {\doibase https://doi.org/10.1016/j.aop.2017.02.006}
  {\bibfield  {journal} {\bibinfo  {journal} {Ann. Phys.}\ }\textbf {\bibinfo
  {volume} {378}},\ \bibinfo {pages} {448 -- 458} (\bibinfo {year}
  {2017})}\BibitemShut {NoStop}%
\bibitem [{\citenamefont {Baksic}\ and\ \citenamefont
  {Ciuti}(2014)}]{Baksic14}%
  \BibitemOpen
  \bibfield  {author} {\bibinfo {author} {\bibfnamefont {A.}~\bibnamefont
  {Baksic}}\ and\ \bibinfo {author} {\bibfnamefont {C.}~\bibnamefont {Ciuti}},\
  }\bibfield  {title} {\enquote {\bibinfo {title} {Controlling discrete and
  continuous symmetries in ``superradiant'' phase transitions with circuit qed
  systems},}\ }\href {\doibase 10.1103/PhysRevLett.112.173601} {\bibfield
  {journal} {\bibinfo  {journal} {Phys. Rev. Lett.}\ }\textbf {\bibinfo
  {volume} {112}},\ \bibinfo {pages} {173601} (\bibinfo {year}
  {2014})}\BibitemShut {NoStop}%
\bibitem [{\citenamefont {Hepp}\ and\ \citenamefont
  {Lieb}(1973{\natexlab{a}})}]{HEPP73}%
  \BibitemOpen
  \bibfield  {author} {\bibinfo {author} {\bibfnamefont {K.}~\bibnamefont
  {Hepp}}\ and\ \bibinfo {author} {\bibfnamefont {E.~H.}\ \bibnamefont
  {Lieb}},\ }\bibfield  {title} {\enquote {\bibinfo {title} {On the
  superradiant phase transition for molecules in a quantized radiation field:
  the dicke maser model},}\ }\href {\doibase
  https://doi.org/10.1016/0003-4916(73)90039-0} {\bibfield  {journal} {\bibinfo
   {journal} {Ann. Phys.}\ }\textbf {\bibinfo {volume} {76}},\ \bibinfo {pages}
  {360 -- 404} (\bibinfo {year} {1973}{\natexlab{a}})}\BibitemShut {NoStop}%
\bibitem [{\citenamefont {Lambert}\ \emph {et~al.}(2004)\citenamefont
  {Lambert}, \citenamefont {Emary},\ and\ \citenamefont
  {Brandes}}]{lambert2004entanglement}%
  \BibitemOpen
  \bibfield  {author} {\bibinfo {author} {\bibfnamefont {N.}~\bibnamefont
  {Lambert}}, \bibinfo {author} {\bibfnamefont {C.}~\bibnamefont {Emary}}, \
  and\ \bibinfo {author} {\bibfnamefont {T.}~\bibnamefont {Brandes}},\
  }\bibfield  {title} {\enquote {\bibinfo {title} {Entanglement and the phase
  transition in single-mode superradiance},}\ }\href
  {https://journals.aps.org/prl/abstract/10.1103/PhysRevLett.92.073602}
  {\bibfield  {journal} {\bibinfo  {journal} {Phys. Rev. Lett.}\ }\textbf
  {\bibinfo {volume} {92}},\ \bibinfo {pages} {073602} (\bibinfo {year}
  {2004})}\BibitemShut {NoStop}%
\bibitem [{\citenamefont {Lee}\ and\ \citenamefont
  {Johnson}(2004)}]{lee2004first}%
  \BibitemOpen
  \bibfield  {author} {\bibinfo {author} {\bibfnamefont {C.~F.}\ \bibnamefont
  {Lee}}\ and\ \bibinfo {author} {\bibfnamefont {N.~F.}\ \bibnamefont
  {Johnson}},\ }\bibfield  {title} {\enquote {\bibinfo {title} {First-order
  superradiant phase transitions in a multiqubit cavity system},}\ }\href
  {https://journals.aps.org/prl/abstract/10.1103/PhysRevLett.93.083001}
  {\bibfield  {journal} {\bibinfo  {journal} {Phys. Rev. Lett.}\ }\textbf
  {\bibinfo {volume} {93}},\ \bibinfo {pages} {083001} (\bibinfo {year}
  {2004})}\BibitemShut {NoStop}%
\bibitem [{\citenamefont {Bu{\v{z}}ek}\ \emph {et~al.}(2005)\citenamefont
  {Bu{\v{z}}ek}, \citenamefont {Orszag},\ and\ \citenamefont
  {Ro{\v{s}}ko}}]{buvzek2005instability}%
  \BibitemOpen
  \bibfield  {author} {\bibinfo {author} {\bibfnamefont {V.}~\bibnamefont
  {Bu{\v{z}}ek}}, \bibinfo {author} {\bibfnamefont {M.}~\bibnamefont {Orszag}},
  \ and\ \bibinfo {author} {\bibfnamefont {M.}~\bibnamefont {Ro{\v{s}}ko}},\
  }\bibfield  {title} {\enquote {\bibinfo {title} {Instability and entanglement
  of the ground state of the dicke model},}\ }\href
  {https://journals.aps.org/prl/abstract/10.1103/PhysRevLett.94.163601}
  {\bibfield  {journal} {\bibinfo  {journal} {Phys. Rev. Lett.}\ }\textbf
  {\bibinfo {volume} {94}},\ \bibinfo {pages} {163601} (\bibinfo {year}
  {2005})}\BibitemShut {NoStop}%
\bibitem [{\citenamefont {Larson}\ and\ \citenamefont
  {Irish}(2017)}]{larson2017some}%
  \BibitemOpen
  \bibfield  {author} {\bibinfo {author} {\bibfnamefont {J.}~\bibnamefont
  {Larson}}\ and\ \bibinfo {author} {\bibfnamefont {E.~K.}\ \bibnamefont
  {Irish}},\ }\bibfield  {title} {\enquote {\bibinfo {title} {Some remarks on
  ‘superradiant’phase transitions in light-matter systems},}\ }\href
  {\doibase 10.1088/1751-8121/aa65dc} {\bibfield  {journal} {\bibinfo
  {journal} {J. Phys. A: Math. Theor.}\ }\textbf {\bibinfo {volume} {50}},\
  \bibinfo {pages} {174002} (\bibinfo {year} {2017})}\BibitemShut {NoStop}%
\bibitem [{\citenamefont {Casanova}\ \emph {et~al.}(2010)\citenamefont
  {Casanova}, \citenamefont {Romero}, \citenamefont {Lizuain}, \citenamefont
  {Garc{\'\i}a-Ripoll},\ and\ \citenamefont {Solano}}]{casanova2010deep}%
  \BibitemOpen
  \bibfield  {author} {\bibinfo {author} {\bibfnamefont {J.}~\bibnamefont
  {Casanova}}, \bibinfo {author} {\bibfnamefont {G.}~\bibnamefont {Romero}},
  \bibinfo {author} {\bibfnamefont {I.}~\bibnamefont {Lizuain}}, \bibinfo
  {author} {\bibfnamefont {J.~J.}\ \bibnamefont {Garc{\'\i}a-Ripoll}}, \ and\
  \bibinfo {author} {\bibfnamefont {E.}~\bibnamefont {Solano}},\ }\bibfield
  {title} {\enquote {\bibinfo {title} {Deep strong coupling regime of the
  jaynes-cummings model},}\ }\href {\doibase 10.1103/PhysRevLett.105.263603}
  {\bibfield  {journal} {\bibinfo  {journal} {Phys. Rev. Lett.}\ }\textbf
  {\bibinfo {volume} {105}},\ \bibinfo {pages} {263603} (\bibinfo {year}
  {2010})}\BibitemShut {NoStop}%
\bibitem [{\citenamefont {Castanos}\ \emph {et~al.}(2009)\citenamefont
  {Castanos}, \citenamefont {L{\'o}pez-Pena}, \citenamefont {Nahmad-Achar},
  \citenamefont {Hirsch}, \citenamefont {L{\'o}pez-Moreno},\ and\ \citenamefont
  {Vitela}}]{castanos2009coherent}%
  \BibitemOpen
  \bibfield  {author} {\bibinfo {author} {\bibfnamefont {O.}~\bibnamefont
  {Castanos}}, \bibinfo {author} {\bibfnamefont {R.}~\bibnamefont
  {L{\'o}pez-Pena}}, \bibinfo {author} {\bibfnamefont {E.}~\bibnamefont
  {Nahmad-Achar}}, \bibinfo {author} {\bibfnamefont {J.~G.}\ \bibnamefont
  {Hirsch}}, \bibinfo {author} {\bibfnamefont {E.}~\bibnamefont
  {L{\'o}pez-Moreno}}, \ and\ \bibinfo {author} {\bibfnamefont {J.~E.}\
  \bibnamefont {Vitela}},\ }\bibfield  {title} {\enquote {\bibinfo {title}
  {Coherent state description of the ground state in the tavis--cummings model
  and its quantum phase transitions},}\ }\href {\doibase
  10.1088/0031-8949/79/06/065405} {\bibfield  {journal} {\bibinfo  {journal}
  {Phys. Scr.}\ }\textbf {\bibinfo {volume} {79}},\ \bibinfo {pages} {065405}
  (\bibinfo {year} {2009})}\BibitemShut {NoStop}%
\bibitem [{\citenamefont {Hepp}\ and\ \citenamefont
  {Lieb}(1973{\natexlab{b}})}]{hepp1973equilibrium}%
  \BibitemOpen
  \bibfield  {author} {\bibinfo {author} {\bibfnamefont {K.}~\bibnamefont
  {Hepp}}\ and\ \bibinfo {author} {\bibfnamefont {E.~H.}\ \bibnamefont
  {Lieb}},\ }\bibfield  {title} {\enquote {\bibinfo {title} {Equilibrium
  statistical mechanics of matter interacting with the quantized radiation
  field},}\ }\href {\doibase 10.1103/PhysRevA.8.2517} {\bibfield  {journal}
  {\bibinfo  {journal} {Phys. Rev. A}\ }\textbf {\bibinfo {volume} {8}},\
  \bibinfo {pages} {2517} (\bibinfo {year} {1973}{\natexlab{b}})}\BibitemShut
  {NoStop}%
\bibitem [{\citenamefont {Soriente}\ \emph {et~al.}(2018)\citenamefont
  {Soriente}, \citenamefont {Donner}, \citenamefont {Chitra},\ and\
  \citenamefont {Zilberberg}}]{soriente2018dissipation}%
  \BibitemOpen
  \bibfield  {author} {\bibinfo {author} {\bibfnamefont {M.}~\bibnamefont
  {Soriente}}, \bibinfo {author} {\bibfnamefont {T.}~\bibnamefont {Donner}},
  \bibinfo {author} {\bibfnamefont {R.}~\bibnamefont {Chitra}}, \ and\ \bibinfo
  {author} {\bibfnamefont {O.}~\bibnamefont {Zilberberg}},\ }\bibfield  {title}
  {\enquote {\bibinfo {title} {Dissipation-induced anomalous multicritical
  phenomena},}\ }\href {\doibase 10.1103/PhysRevLett.120.183603} {\bibfield
  {journal} {\bibinfo  {journal} {Phys. Rev. Lett.}\ }\textbf {\bibinfo
  {volume} {120}},\ \bibinfo {pages} {183603} (\bibinfo {year}
  {2018})}\BibitemShut {NoStop}%
\bibitem [{\citenamefont {Gu}\ \emph {et~al.}(2017)\citenamefont {Gu},
  \citenamefont {Kockum}, \citenamefont {Miranowicz}, \citenamefont {Liu},\
  and\ \citenamefont {Nori}}]{gu2017microwave}%
  \BibitemOpen
  \bibfield  {author} {\bibinfo {author} {\bibfnamefont {X.}~\bibnamefont
  {Gu}}, \bibinfo {author} {\bibfnamefont {A.~F.}\ \bibnamefont {Kockum}},
  \bibinfo {author} {\bibfnamefont {A.}~\bibnamefont {Miranowicz}}, \bibinfo
  {author} {\bibfnamefont {Y.~X.}\ \bibnamefont {Liu}}, \ and\ \bibinfo
  {author} {\bibfnamefont {F.}~\bibnamefont {Nori}},\ }\bibfield  {title}
  {\enquote {\bibinfo {title} {Microwave photonics with superconducting quantum
  circuits},}\ }\href
  {https://www.sciencedirect.com/science/article/pii/S0370157317303290}
  {\bibfield  {journal} {\bibinfo  {journal} {Phys. Rep.}\ }\textbf {\bibinfo
  {volume} {718}},\ \bibinfo {pages} {1--102} (\bibinfo {year}
  {2017})}\BibitemShut {NoStop}%
\bibitem [{\citenamefont {Armata}\ \emph {et~al.}(2017)\citenamefont {Armata},
  \citenamefont {Calajo}, \citenamefont {Jaako}, \citenamefont {Kim},\ and\
  \citenamefont {Rabl}}]{armata2017harvesting}%
  \BibitemOpen
  \bibfield  {author} {\bibinfo {author} {\bibfnamefont {F.}~\bibnamefont
  {Armata}}, \bibinfo {author} {\bibfnamefont {G.}~\bibnamefont {Calajo}},
  \bibinfo {author} {\bibfnamefont {T.}~\bibnamefont {Jaako}}, \bibinfo
  {author} {\bibfnamefont {M.~S.}\ \bibnamefont {Kim}}, \ and\ \bibinfo
  {author} {\bibfnamefont {P.}~\bibnamefont {Rabl}},\ }\bibfield  {title}
  {\enquote {\bibinfo {title} {Harvesting multiqubit entanglement from
  ultrastrong interactions in circuit quantum electrodynamics},}\ }\href
  {https://journals.aps.org/prl/abstract/10.1103/PhysRevLett.119.183602}
  {\bibfield  {journal} {\bibinfo  {journal} {Phys. Rev. Lett.}\ }\textbf
  {\bibinfo {volume} {119}},\ \bibinfo {pages} {183602} (\bibinfo {year}
  {2017})}\BibitemShut {NoStop}%
\bibitem [{\citenamefont {Niemczyk}\ \emph {et~al.}(2010)\citenamefont
  {Niemczyk}, \citenamefont {Deppe}, \citenamefont {Huebl}, \citenamefont
  {Menzel}, \citenamefont {Hocke}, \citenamefont {Schwarz}, \citenamefont
  {Garcia-Ripoll}, \citenamefont {Zueco}, \citenamefont {H{\"u}mmer},
  \citenamefont {Solano}, \citenamefont {Marx},\ and\ \citenamefont
  {Gross}}]{niemczyk2010circuit}%
  \BibitemOpen
  \bibfield  {author} {\bibinfo {author} {\bibfnamefont {T.}~\bibnamefont
  {Niemczyk}}, \bibinfo {author} {\bibfnamefont {F.}~\bibnamefont {Deppe}},
  \bibinfo {author} {\bibfnamefont {H.}~\bibnamefont {Huebl}}, \bibinfo
  {author} {\bibfnamefont {E.~P.}\ \bibnamefont {Menzel}}, \bibinfo {author}
  {\bibfnamefont {F.}~\bibnamefont {Hocke}}, \bibinfo {author} {\bibfnamefont
  {M.~J.}\ \bibnamefont {Schwarz}}, \bibinfo {author} {\bibfnamefont {J.~J.}\
  \bibnamefont {Garcia-Ripoll}}, \bibinfo {author} {\bibfnamefont
  {D.}~\bibnamefont {Zueco}}, \bibinfo {author} {\bibfnamefont
  {T.}~\bibnamefont {H{\"u}mmer}}, \bibinfo {author} {\bibfnamefont
  {E.}~\bibnamefont {Solano}}, \bibinfo {author} {\bibfnamefont
  {A.}~\bibnamefont {Marx}}, \ and\ \bibinfo {author} {\bibfnamefont
  {R.}~\bibnamefont {Gross}},\ }\bibfield  {title} {\enquote {\bibinfo {title}
  {Circuit quantum electrodynamics in the ultrastrong-coupling regime},}\
  }\href {https://www.nature.com/articles/nphys1730} {\bibfield  {journal}
  {\bibinfo  {journal} {Nat. Phys.}\ }\textbf {\bibinfo {volume} {6}},\
  \bibinfo {pages} {772} (\bibinfo {year} {2010})}\BibitemShut {NoStop}%
\bibitem [{\citenamefont {Hioe}(1973)}]{hioe1973phase}%
  \BibitemOpen
  \bibfield  {author} {\bibinfo {author} {\bibfnamefont {F.~T.}\ \bibnamefont
  {Hioe}},\ }\bibfield  {title} {\enquote {\bibinfo {title} {Phase transitions
  in some generalized dicke models of superradiance},}\ }\href {\doibase
  10.1103/PhysRevA.8.1440} {\bibfield  {journal} {\bibinfo  {journal} {Phys.
  Rev. A}\ }\textbf {\bibinfo {volume} {8}},\ \bibinfo {pages} {1440} (\bibinfo
  {year} {1973})}\BibitemShut {NoStop}%
\bibitem [{\citenamefont {Bakemeier}\ \emph {et~al.}(2012)\citenamefont
  {Bakemeier}, \citenamefont {Alvermann},\ and\ \citenamefont
  {Fehske}}]{bakemeier2012quantum}%
  \BibitemOpen
  \bibfield  {author} {\bibinfo {author} {\bibfnamefont {L.}~\bibnamefont
  {Bakemeier}}, \bibinfo {author} {\bibfnamefont {A.}~\bibnamefont
  {Alvermann}}, \ and\ \bibinfo {author} {\bibfnamefont {H.}~\bibnamefont
  {Fehske}},\ }\bibfield  {title} {\enquote {\bibinfo {title} {Quantum phase
  transition in the dicke model with critical and noncritical entanglement},}\
  }\href {\doibase 10.1103/PhysRevA.85.043821} {\bibfield  {journal} {\bibinfo
  {journal} {Phys. Rev. A}\ }\textbf {\bibinfo {volume} {85}},\ \bibinfo
  {pages} {043821} (\bibinfo {year} {2012})}\BibitemShut {NoStop}%
\bibitem [{\citenamefont {Guti{\'e}rrez-J{\'a}uregui}\ and\ \citenamefont
  {Carmichael}(2018)}]{gutierrez2018dissipative}%
  \BibitemOpen
  \bibfield  {author} {\bibinfo {author} {\bibfnamefont {R.}~\bibnamefont
  {Guti{\'e}rrez-J{\'a}uregui}}\ and\ \bibinfo {author} {\bibfnamefont {H.~J.}\
  \bibnamefont {Carmichael}},\ }\bibfield  {title} {\enquote {\bibinfo {title}
  {Dissipative quantum phase transitions of light in a generalized
  jaynes-cummings-rabi model},}\ }\href {\doibase 10.1103/PhysRevA.98.023804}
  {\bibfield  {journal} {\bibinfo  {journal} {Phys. Rev. A}\ }\textbf {\bibinfo
  {volume} {98}},\ \bibinfo {pages} {023804} (\bibinfo {year}
  {2018})}\BibitemShut {NoStop}%
\bibitem [{\citenamefont {Wang}\ and\ \citenamefont {Hioe}(1973)}]{Wang73}%
  \BibitemOpen
  \bibfield  {author} {\bibinfo {author} {\bibfnamefont {Y.~K.}\ \bibnamefont
  {Wang}}\ and\ \bibinfo {author} {\bibfnamefont {F.~T.}\ \bibnamefont
  {Hioe}},\ }\bibfield  {title} {\enquote {\bibinfo {title} {Phase transition
  in the dicke model of superradiance},}\ }\href {\doibase
  10.1103/PhysRevA.7.831} {\bibfield  {journal} {\bibinfo  {journal} {Phys.
  Rev. A}\ }\textbf {\bibinfo {volume} {7}},\ \bibinfo {pages} {831--836}
  (\bibinfo {year} {1973})}\BibitemShut {NoStop}%
\bibitem [{\citenamefont {Ashhab}(2013)}]{ashhab2013superradiance}%
  \BibitemOpen
  \bibfield  {author} {\bibinfo {author} {\bibfnamefont {S.}~\bibnamefont
  {Ashhab}},\ }\bibfield  {title} {\enquote {\bibinfo {title} {Superradiance
  transition in a system with a single qubit and a single oscillator},}\ }\href
  {\doibase 10.1103/PhysRevA.87.013826} {\bibfield  {journal} {\bibinfo
  {journal} {Phys. Rev. A}\ }\textbf {\bibinfo {volume} {87}},\ \bibinfo
  {pages} {013826} (\bibinfo {year} {2013})}\BibitemShut {NoStop}%
\bibitem [{\citenamefont {Hwang}\ \emph {et~al.}(2015)\citenamefont {Hwang},
  \citenamefont {Puebla},\ and\ \citenamefont {Plenio}}]{hwang2015quantum}%
  \BibitemOpen
  \bibfield  {author} {\bibinfo {author} {\bibfnamefont {M.~J.}\ \bibnamefont
  {Hwang}}, \bibinfo {author} {\bibfnamefont {R.}~\bibnamefont {Puebla}}, \
  and\ \bibinfo {author} {\bibfnamefont {M.~B.}\ \bibnamefont {Plenio}},\
  }\bibfield  {title} {\enquote {\bibinfo {title} {Quantum phase transition and
  universal dynamics in the rabi model},}\ }\href {\doibase
  10.1103/PhysRevLett.115.180404} {\bibfield  {journal} {\bibinfo  {journal}
  {Phys. Rev. Lett.}\ }\textbf {\bibinfo {volume} {115}},\ \bibinfo {pages}
  {180404} (\bibinfo {year} {2015})}\BibitemShut {NoStop}%
\bibitem [{\citenamefont {Hwang}\ and\ \citenamefont
  {Plenio}(2016)}]{hwang2016quantum}%
  \BibitemOpen
  \bibfield  {author} {\bibinfo {author} {\bibfnamefont {M.~J.}\ \bibnamefont
  {Hwang}}\ and\ \bibinfo {author} {\bibfnamefont {M.~B.}\ \bibnamefont
  {Plenio}},\ }\bibfield  {title} {\enquote {\bibinfo {title} {Quantum phase
  transition in the finite jaynes-cummings lattice systems},}\ }\href {\doibase
  10.1103/PhysRevLett.117.123602} {\bibfield  {journal} {\bibinfo  {journal}
  {Phys. Rev. Lett.}\ }\textbf {\bibinfo {volume} {117}},\ \bibinfo {pages}
  {123602} (\bibinfo {year} {2016})}\BibitemShut {NoStop}%
\bibitem [{\citenamefont {Zou}\ \emph {et~al.}(2013)\citenamefont {Zou},
  \citenamefont {Liu}, \citenamefont {Feng}, \citenamefont {Yang},
  \citenamefont {Chen},\ and\ \citenamefont {Twamley}}]{zou2013quantum}%
  \BibitemOpen
  \bibfield  {author} {\bibinfo {author} {\bibfnamefont {J.~H.}\ \bibnamefont
  {Zou}}, \bibinfo {author} {\bibfnamefont {T.}~\bibnamefont {Liu}}, \bibinfo
  {author} {\bibfnamefont {M.}~\bibnamefont {Feng}}, \bibinfo {author}
  {\bibfnamefont {W.~L.}\ \bibnamefont {Yang}}, \bibinfo {author}
  {\bibfnamefont {C.~Y.}\ \bibnamefont {Chen}}, \ and\ \bibinfo {author}
  {\bibfnamefont {J.}~\bibnamefont {Twamley}},\ }\bibfield  {title} {\enquote
  {\bibinfo {title} {Quantum phase transition in a driven tavis--cummings
  model},}\ }\href {\doibase 10.1088/1367-2630/15/12/123032} {\bibfield
  {journal} {\bibinfo  {journal} {New J. Phys.}\ }\textbf {\bibinfo {volume}
  {15}},\ \bibinfo {pages} {123032} (\bibinfo {year} {2013})}\BibitemShut
  {NoStop}%
\bibitem [{\citenamefont {Narducci}\ \emph {et~al.}(1978)\citenamefont
  {Narducci}, \citenamefont {Feng}, \citenamefont {Gilmore},\ and\
  \citenamefont {Agarwal}}]{narducci1978transient}%
  \BibitemOpen
  \bibfield  {author} {\bibinfo {author} {\bibfnamefont {L.~M.}\ \bibnamefont
  {Narducci}}, \bibinfo {author} {\bibfnamefont {D.~H.}\ \bibnamefont {Feng}},
  \bibinfo {author} {\bibfnamefont {R.}~\bibnamefont {Gilmore}}, \ and\
  \bibinfo {author} {\bibfnamefont {G.~S.}\ \bibnamefont {Agarwal}},\
  }\bibfield  {title} {\enquote {\bibinfo {title} {Transient and steady-state
  behavior of collective atomic systems driven by a classical field},}\ }\href
  {https://journals.aps.org/pra/abstract/10.1103/PhysRevA.18.1571} {\bibfield
  {journal} {\bibinfo  {journal} {Phys. Rev. A}\ }\textbf {\bibinfo {volume}
  {18}},\ \bibinfo {pages} {1571} (\bibinfo {year} {1978})}\BibitemShut
  {NoStop}%
\bibitem [{\citenamefont {Puri}\ and\ \citenamefont
  {Lawande}(1980)}]{puri1980exact}%
  \BibitemOpen
  \bibfield  {author} {\bibinfo {author} {\bibfnamefont {R.~R.}\ \bibnamefont
  {Puri}}\ and\ \bibinfo {author} {\bibfnamefont {S.~V.}\ \bibnamefont
  {Lawande}},\ }\bibfield  {title} {\enquote {\bibinfo {title} {Exact
  thermodynamic behaviour of a collective atomic system in an external
  field},}\ }\href
  {https://www.sciencedirect.com/science/article/pii/0378437180901983}
  {\bibfield  {journal} {\bibinfo  {journal} {Phys. A, Stat. Mech. Appl.}\
  }\textbf {\bibinfo {volume} {101}},\ \bibinfo {pages} {599--612} (\bibinfo
  {year} {1980})}\BibitemShut {NoStop}%
\bibitem [{\citenamefont {Schneider}\ and\ \citenamefont
  {Milburn}(2002)}]{schneider2002entanglement}%
  \BibitemOpen
  \bibfield  {author} {\bibinfo {author} {\bibfnamefont {S.}~\bibnamefont
  {Schneider}}\ and\ \bibinfo {author} {\bibfnamefont {G.~J.}\ \bibnamefont
  {Milburn}},\ }\bibfield  {title} {\enquote {\bibinfo {title} {Entanglement in
  the steady state of a collective-angular-momentum (dicke) model},}\ }\href
  {https://journals.aps.org/pra/abstract/10.1103/PhysRevA.65.042107} {\bibfield
   {journal} {\bibinfo  {journal} {Phys. Rev. A}\ }\textbf {\bibinfo {volume}
  {65}},\ \bibinfo {pages} {042107} (\bibinfo {year} {2002})}\BibitemShut
  {NoStop}%
\bibitem [{\citenamefont {Minganti}\ \emph {et~al.}(2018)\citenamefont
  {Minganti}, \citenamefont {Biella}, \citenamefont {Bartolo},\ and\
  \citenamefont {Ciuti}}]{minganti2018spectral}%
  \BibitemOpen
  \bibfield  {author} {\bibinfo {author} {\bibfnamefont {F.}~\bibnamefont
  {Minganti}}, \bibinfo {author} {\bibfnamefont {A.}~\bibnamefont {Biella}},
  \bibinfo {author} {\bibfnamefont {N.}~\bibnamefont {Bartolo}}, \ and\
  \bibinfo {author} {\bibfnamefont {C.}~\bibnamefont {Ciuti}},\ }\bibfield
  {title} {\enquote {\bibinfo {title} {Spectral theory of liouvillians for
  dissipative phase transitions},}\ }\href
  {https://journals.aps.org/pra/abstract/10.1103/PhysRevA.98.042118} {\bibfield
   {journal} {\bibinfo  {journal} {Phys. Rev. A}\ }\textbf {\bibinfo {volume}
  {98}},\ \bibinfo {pages} {042118} (\bibinfo {year} {2018})}\BibitemShut
  {NoStop}%
\bibitem [{\citenamefont {Casteels}\ \emph {et~al.}(2017)\citenamefont
  {Casteels}, \citenamefont {Fazio},\ and\ \citenamefont
  {Ciuti}}]{casteels2017critical}%
  \BibitemOpen
  \bibfield  {author} {\bibinfo {author} {\bibfnamefont {W.}~\bibnamefont
  {Casteels}}, \bibinfo {author} {\bibfnamefont {R.}~\bibnamefont {Fazio}}, \
  and\ \bibinfo {author} {\bibfnamefont {C.}~\bibnamefont {Ciuti}},\ }\bibfield
   {title} {\enquote {\bibinfo {title} {Critical dynamical properties of a
  first-order dissipative phase transition},}\ }\href
  {https://journals.aps.org/pra/abstract/10.1103/PhysRevA.95.012128} {\bibfield
   {journal} {\bibinfo  {journal} {Phys. Rev. A}\ }\textbf {\bibinfo {volume}
  {95}},\ \bibinfo {pages} {012128} (\bibinfo {year} {2017})}\BibitemShut
  {NoStop}%
\bibitem [{\citenamefont {Carmichael}(2015)}]{carmichael2015breakdown}%
  \BibitemOpen
  \bibfield  {author} {\bibinfo {author} {\bibfnamefont {H.~J.}\ \bibnamefont
  {Carmichael}},\ }\bibfield  {title} {\enquote {\bibinfo {title} {Breakdown of
  photon blockade: A dissipative quantum phase transition in zero
  dimensions},}\ }\href
  {https://journals.aps.org/prx/abstract/10.1103/PhysRevX.5.031028} {\bibfield
  {journal} {\bibinfo  {journal} {Phys. Rev. X}\ }\textbf {\bibinfo {volume}
  {5}},\ \bibinfo {pages} {031028} (\bibinfo {year} {2015})}\BibitemShut
  {NoStop}%
\bibitem [{\citenamefont {Huang}\ and\ \citenamefont
  {Agarwal}(2009)}]{huang2009normal}%
  \BibitemOpen
  \bibfield  {author} {\bibinfo {author} {\bibfnamefont {S.~M.}\ \bibnamefont
  {Huang}}\ and\ \bibinfo {author} {\bibfnamefont {G.~S.}\ \bibnamefont
  {Agarwal}},\ }\bibfield  {title} {\enquote {\bibinfo {title} {Normal-mode
  splitting in a coupled system of a nanomechanical oscillator and a parametric
  amplifier cavity},}\ }\href
  {https://journals.aps.org/pra/abstract/10.1103/PhysRevA.80.033807} {\bibfield
   {journal} {\bibinfo  {journal} {Phys. Rev. A}\ }\textbf {\bibinfo {volume}
  {80}},\ \bibinfo {pages} {033807} (\bibinfo {year} {2009})}\BibitemShut
  {NoStop}%
\bibitem [{\citenamefont {Leroux}\ \emph {et~al.}(2018)\citenamefont {Leroux},
  \citenamefont {Govia},\ and\ \citenamefont {Clerk}}]{leroux2018enhancing}%
  \BibitemOpen
  \bibfield  {author} {\bibinfo {author} {\bibfnamefont {C.}~\bibnamefont
  {Leroux}}, \bibinfo {author} {\bibfnamefont {L.~C.~G.}\ \bibnamefont
  {Govia}}, \ and\ \bibinfo {author} {\bibfnamefont {A.~A.}\ \bibnamefont
  {Clerk}},\ }\bibfield  {title} {\enquote {\bibinfo {title} {Enhancing cavity
  quantum electrodynamics via antisqueezing: synthetic ultrastrong coupling},}\
  }\href {https://journals.aps.org/prl/abstract/10.1103/PhysRevLett.120.093602}
  {\bibfield  {journal} {\bibinfo  {journal} {Phys. Rev. Lett.}\ }\textbf
  {\bibinfo {volume} {120}},\ \bibinfo {pages} {093602} (\bibinfo {year}
  {2018})}\BibitemShut {NoStop}%
\bibitem [{\citenamefont {M{\"u}ller}\ \emph {et~al.}(2012)\citenamefont
  {M{\"u}ller}, \citenamefont {Diehl}, \citenamefont {Pupillo},\ and\
  \citenamefont {Zoller}}]{muller2012engineered}%
  \BibitemOpen
  \bibfield  {author} {\bibinfo {author} {\bibfnamefont {M.}~\bibnamefont
  {M{\"u}ller}}, \bibinfo {author} {\bibfnamefont {S.}~\bibnamefont {Diehl}},
  \bibinfo {author} {\bibfnamefont {G.}~\bibnamefont {Pupillo}}, \ and\
  \bibinfo {author} {\bibfnamefont {P.}~\bibnamefont {Zoller}},\ }\bibfield
  {title} {\enquote {\bibinfo {title} {Engineered open systems and quantum
  simulations with atoms and ions},}\ }in\ \href
  {https://doi.org/10.1016/B978-0-12-396482-3.00001-6} {\emph {\bibinfo
  {booktitle} {Adv. At., Mol., Opt. Phys.}}},\ Vol.~\bibinfo {volume} {61}\
  (\bibinfo  {publisher} {Elsevier},\ \bibinfo {year} {2012})\ pp.\ \bibinfo
  {pages} {1--80}\BibitemShut {NoStop}%
\bibitem [{\citenamefont {Bernien}\ \emph {et~al.}(2017)\citenamefont
  {Bernien}, \citenamefont {Schwartz}, \citenamefont {Keesling}, \citenamefont
  {Levine}, \citenamefont {Omran}, \citenamefont {Pichler}, \citenamefont
  {Choi}, \citenamefont {Zibrov}, \citenamefont {Endres}, \citenamefont
  {Greiner}, \citenamefont {Vuletic},\ and\ \citenamefont
  {Lukin}}]{bernien2017probing}%
  \BibitemOpen
  \bibfield  {author} {\bibinfo {author} {\bibfnamefont {H.}~\bibnamefont
  {Bernien}}, \bibinfo {author} {\bibfnamefont {S.}~\bibnamefont {Schwartz}},
  \bibinfo {author} {\bibfnamefont {A.}~\bibnamefont {Keesling}}, \bibinfo
  {author} {\bibfnamefont {H.}~\bibnamefont {Levine}}, \bibinfo {author}
  {\bibfnamefont {A.}~\bibnamefont {Omran}}, \bibinfo {author} {\bibfnamefont
  {H.}~\bibnamefont {Pichler}}, \bibinfo {author} {\bibfnamefont
  {S.}~\bibnamefont {Choi}}, \bibinfo {author} {\bibfnamefont {A.~S.}\
  \bibnamefont {Zibrov}}, \bibinfo {author} {\bibfnamefont {M.}~\bibnamefont
  {Endres}}, \bibinfo {author} {\bibfnamefont {M.}~\bibnamefont {Greiner}},
  \bibinfo {author} {\bibfnamefont {V.}~\bibnamefont {Vuletic}}, \ and\
  \bibinfo {author} {\bibfnamefont {M.~D.}\ \bibnamefont {Lukin}},\ }\bibfield
  {title} {\enquote {\bibinfo {title} {Probing many-body dynamics on a 51-atom
  quantum simulator},}\ }\href {https://www.nature.com/articles/nature24622}
  {\bibfield  {journal} {\bibinfo  {journal} {Nature}\ }\textbf {\bibinfo
  {volume} {551}},\ \bibinfo {pages} {579} (\bibinfo {year}
  {2017})}\BibitemShut {NoStop}%
\bibitem [{\citenamefont {Ying}\ \emph {et~al.}(2014)\citenamefont {Ying},
  \citenamefont {Lai},\ and\ \citenamefont {Grebogi}}]{ying2014quantum}%
  \BibitemOpen
  \bibfield  {author} {\bibinfo {author} {\bibfnamefont {L.}~\bibnamefont
  {Ying}}, \bibinfo {author} {\bibfnamefont {Y.~C.}\ \bibnamefont {Lai}}, \
  and\ \bibinfo {author} {\bibfnamefont {C.}~\bibnamefont {Grebogi}},\
  }\bibfield  {title} {\enquote {\bibinfo {title} {Quantum manifestation of a
  synchronization transition in optomechanical systems},}\ }\href
  {https://journals.aps.org/pra/abstract/10.1103/PhysRevA.90.053810} {\bibfield
   {journal} {\bibinfo  {journal} {Phys. Rev. A}\ }\textbf {\bibinfo {volume}
  {90}},\ \bibinfo {pages} {053810} (\bibinfo {year} {2014})}\BibitemShut
  {NoStop}%
\bibitem [{\citenamefont {Sieberer}\ \emph {et~al.}(2013)\citenamefont
  {Sieberer}, \citenamefont {Huber}, \citenamefont {Altman},\ and\
  \citenamefont {Diehl}}]{sieberer2013dynamical}%
  \BibitemOpen
  \bibfield  {author} {\bibinfo {author} {\bibfnamefont {L.~M.}\ \bibnamefont
  {Sieberer}}, \bibinfo {author} {\bibfnamefont {S.~D.}\ \bibnamefont {Huber}},
  \bibinfo {author} {\bibfnamefont {E.}~\bibnamefont {Altman}}, \ and\ \bibinfo
  {author} {\bibfnamefont {S.}~\bibnamefont {Diehl}},\ }\bibfield  {title}
  {\enquote {\bibinfo {title} {Dynamical critical phenomena in
  driven-dissipative systems},}\ }\href
  {https://journals.aps.org/prl/abstract/10.1103/PhysRevLett.110.195301}
  {\bibfield  {journal} {\bibinfo  {journal} {Phys. Rev. Lett.}\ }\textbf
  {\bibinfo {volume} {110}},\ \bibinfo {pages} {195301} (\bibinfo {year}
  {2013})}\BibitemShut {NoStop}%
\bibitem [{\citenamefont {Houck}\ \emph {et~al.}(2012)\citenamefont {Houck},
  \citenamefont {T{\"u}reci},\ and\ \citenamefont {Koch}}]{houck2012chip}%
  \BibitemOpen
  \bibfield  {author} {\bibinfo {author} {\bibfnamefont {A.~A.}\ \bibnamefont
  {Houck}}, \bibinfo {author} {\bibfnamefont {H.~E.}\ \bibnamefont
  {T{\"u}reci}}, \ and\ \bibinfo {author} {\bibfnamefont {J.}~\bibnamefont
  {Koch}},\ }\bibfield  {title} {\enquote {\bibinfo {title} {On-chip quantum
  simulation with superconducting circuits},}\ }\href
  {https://www.nature.com/articles/nphys2251} {\bibfield  {journal} {\bibinfo
  {journal} {Nat. Phys.}\ }\textbf {\bibinfo {volume} {8}},\ \bibinfo {pages}
  {292} (\bibinfo {year} {2012})}\BibitemShut {NoStop}%
\bibitem [{\citenamefont {Fitzpatrick}\ \emph {et~al.}(2017)\citenamefont
  {Fitzpatrick}, \citenamefont {Sundaresan}, \citenamefont {Li}, \citenamefont
  {Koch},\ and\ \citenamefont {Houck}}]{fitzpatrick2017observation}%
  \BibitemOpen
  \bibfield  {author} {\bibinfo {author} {\bibfnamefont {M.}~\bibnamefont
  {Fitzpatrick}}, \bibinfo {author} {\bibfnamefont {N.~M.}\ \bibnamefont
  {Sundaresan}}, \bibinfo {author} {\bibfnamefont {A.~C.~Y.}\ \bibnamefont
  {Li}}, \bibinfo {author} {\bibfnamefont {J.}~\bibnamefont {Koch}}, \ and\
  \bibinfo {author} {\bibfnamefont {A.~A.}\ \bibnamefont {Houck}},\ }\bibfield
  {title} {\enquote {\bibinfo {title} {Observation of a dissipative phase
  transition in a one-dimensional circuit qed lattice},}\ }\href
  {https://journals.aps.org/prx/abstract/10.1103/PhysRevX.7.011016} {\bibfield
  {journal} {\bibinfo  {journal} {Phys. Rev. X}\ }\textbf {\bibinfo {volume}
  {7}},\ \bibinfo {pages} {011016} (\bibinfo {year} {2017})}\BibitemShut
  {NoStop}%
\bibitem [{\citenamefont {Bhaseen}\ \emph {et~al.}(2012)\citenamefont
  {Bhaseen}, \citenamefont {Mayoh}, \citenamefont {Simons},\ and\ \citenamefont
  {Keeling}}]{PhysRevA.85.013817}%
  \BibitemOpen
  \bibfield  {author} {\bibinfo {author} {\bibfnamefont {M.~J.}\ \bibnamefont
  {Bhaseen}}, \bibinfo {author} {\bibfnamefont {J.}~\bibnamefont {Mayoh}},
  \bibinfo {author} {\bibfnamefont {B.~D.}\ \bibnamefont {Simons}}, \ and\
  \bibinfo {author} {\bibfnamefont {J.}~\bibnamefont {Keeling}},\ }\bibfield
  {title} {\enquote {\bibinfo {title} {Dynamics of nonequilibrium dicke
  models},}\ }\href {\doibase 10.1103/PhysRevA.85.013817} {\bibfield  {journal}
  {\bibinfo  {journal} {Phys. Rev. A}\ }\textbf {\bibinfo {volume} {85}},\
  \bibinfo {pages} {013817} (\bibinfo {year} {2012})}\BibitemShut {NoStop}%
\bibitem [{\citenamefont {Agarwal}(2012)}]{agarwal2012quantum}%
  \BibitemOpen
  \bibfield  {author} {\bibinfo {author} {\bibfnamefont {G.~S.}\ \bibnamefont
  {Agarwal}},\ }\href@noop {} {\emph {\bibinfo {title} {Quantum optics}}}\
  (\bibinfo  {publisher} {Cambridge University Press},\ \bibinfo {year}
  {2012})\BibitemShut {NoStop}%
\bibitem [{\citenamefont {Wigner}(1932)}]{PhysRev.40.749}%
  \BibitemOpen
  \bibfield  {author} {\bibinfo {author} {\bibfnamefont {E.}~\bibnamefont
  {Wigner}},\ }\bibfield  {title} {\enquote {\bibinfo {title} {On the quantum
  correction for thermodynamic equilibrium},}\ }\href {\doibase
  10.1103/PhysRev.40.749} {\bibfield  {journal} {\bibinfo  {journal} {Phys.
  Rev.}\ }\textbf {\bibinfo {volume} {40}},\ \bibinfo {pages} {749--759}
  (\bibinfo {year} {1932})}\BibitemShut {NoStop}%
\end{thebibliography}%

\end{document}